# Synthesis of DNA Templated Tri-functional Electrically Conducting, Optical and Magnetic nano-chain of $Ni_{core}$-$Au_{shell}$ for Bio-device

Madhuri Mandal*, Kalyan Mandal

*Material Sciences Division, S. N. Bose National Centre for Basic Sciences,
Block JD, Sector III, Salt Lake, Kolkata 700 098, India*
*e-mail : madhuri@bose.res.in

**Abstract:** Synthesis of tri-functional e.g., electrically conducting, optical and magnetic nano-chain of $Ni_{core}$-$Au_{shell}$ have been discussed here. Our Investigation indicates that such material attached with biomolecule 'DNA chain' will have great potentiality in medical instrument and bio computer device.

DNA has well-defined & predictable geometry and diverse & programmable hybridization properties. Therefore various DNA multi assemblies; for example, nanoscaled tiles, cubes, and machines, have been fabricated.[1,2,3] On the other hand biological systems with very small size are very active in a very small scale. Manufacture of various small substances with biological molecules may do many kinds of marvelous things, all on a very small scale. So proper engineering with such system having the property of magnetic as well as optical, have possible applications in brain research, neuro-computation, prosthetics, biosensors, bio-computer etc. For these we need to understand how DNA works with attachment of tiny magnetic and optical materials or nanoparticles.

Now a days the creation of three-dimensional, ordered, crystalline structures of metal nanoparticles guiding by DNA is a big challenge to the researchers.[4,5,6] Mirkin and co-workers[7] described a method of assembling colloidal gold nanoparticles into macroscopic aggregates using DNA as the linking element. Nanowires of noble metals like gold,[7] silver,[8] palladium,[9] platinum,[10] and copper[11] have been metallized on DNA. Synthesis methods, however, required long processing times and high temperatures with multiple steps. People have also synthesized conductive gold on DNA scaffold.[12] Synthesis of DNA templated chain like magnetic and optical material will provide new break through in the nanotechnology research. We have synthesized wire of gold coated Ni magnetic nanoparticles by DNA directing method. For the first time we have reported synthesis of this kind of material. Our material has magnetic, conducting and optical properties all together. This kind of material can be monitored by its optical property in one hand and by magnetic property on the other. Application of this kind of material will also be in versatile fields such as in computer devise, drug delivery, hyperthermia treatment, SERS effect, biosensor etc.

All the reagents used were 99.9% pure and DNAse, RNAse free ultra-pure water was used through out the synthesis method. The aqueous DNA solution has an absorption band at ~260 nm (curve a, Figure 1). The mixing of Ni-sulfate with DNA (curve b, Figure 1) resulted no significant decrease and red shift of absorption peaks for DNA (at ~260 nm) which indicate no aggregation of DNA strands. A slight shift and very small amount of increase of absorbance value of DNA at about 260 nm is observed. But this small shift does not signify any complex formation as there is no considerable shift of the absorption maximum for DNA at 260 nm. After addition of Hydrazene Hydrate and a negligibly small amount of sodium borohydride the solution turned black and this solution show no characteristic absorbance in UV visible spectrum, shown in Figure 1 'c' curve. The $HAuCl_4$ solution in 1/3 molar ratio of Ni-sulfate was added to it. After addition of $HAuCl_4$ the solution turned to blackish pink color with appearance of an additional hump at ~540 nm (Figure 1 d) due to the surface plasmon resonance (SPR) mode of gold nanoparticles. After a long period of time a broad nature SPR peak is observed in curve e, Figure 1, which indicates the formation of thicker gold coating on Ni nanoparticles.[13]

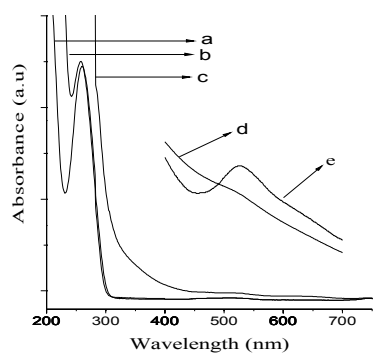

**Figure 1.** UV visible spectra for DNA solution (a), $NiSO_4$ added to DNA solution (b), Ni-nanoparticles attached DNA (c), after addition of $HAuCl_4$ (d), after complete formation of $Ni_{core}Au_{shell}$-DNA composite

TEM images after synthesis of Ni nanoparticles attached on DNA chain and forming the shell of gold on top of Ni nanoparticles are shown in Figure 2 'a' and 'b' respectively. The Ni attached DNA chain is of smaller width of ~40 nm but after gold shell formation the width become larger of ~65 nm. DNA consisting of phosphate, amino groups are good binding agents of metal like Ni, which direct the formation of chain like composite structure.

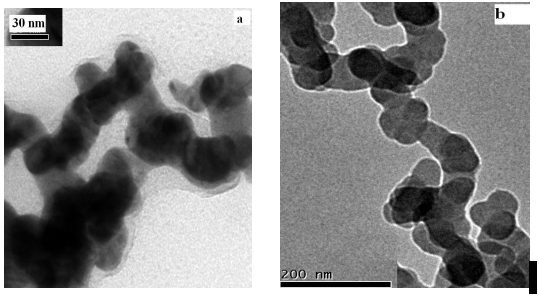

**Figure 2.** TEM images of Ni-nanoparticles attached on DNA (a), and $Ni_{core}Au_{shell}$-DNA (b)

I-V measurement of gold coated Ni-DNA sample shows it can conduct the electricity. The behavior is Ohmic with low resistance. No hysteresis indicates good contacts, continuous structure of one to one nanoparticles attached on DNA.

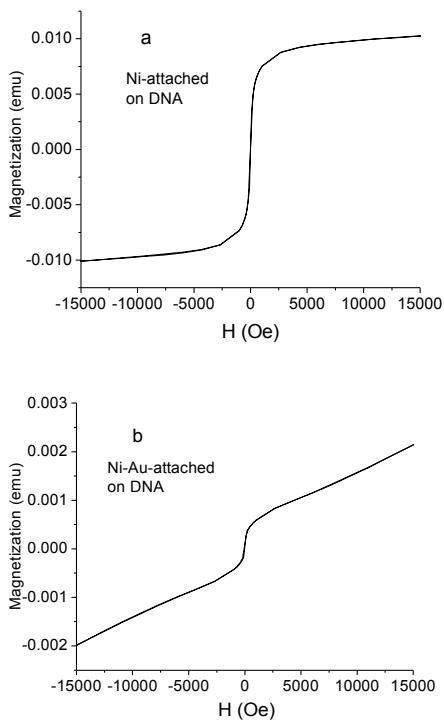

**Figure 3.** Magnetic hysteresis loop of Ni-nanoparticles attached on DNA (a), and $Ni_{core}Au_{shell}$-DNA (b)

Magnetic hysteresis loops for Ni nanoparticles attached on DNA and after its coating by gold are shown in Figure 3 'a' and 'b'. ~0.003 gm of sample was taken for magnetic measurement. Magnetic hysteresis loop indicate super-paramagnetic nature of Ni-attached DNA chain. Curve 'b' indicates both the superpara and paramagnetic character in one hysteresis loop. The superparamagnetic character is due to smaller Ni nanoparticles present in core of the particles and paramagnetic is due to presence of gold as shell. This nature of hysteresis loop again indicates the formation of $Ni_{core}$-$Au_{shell}$ type particles not of alloy type.

People have synthesized electrically conductive gold nanowires using DNA as template exploiting microwave heating method where DNA serves as a reducing and nonspecific capping agent for the growth of nanowires.[12] They have synthesized this kind of material as it has great potential to be used as interconnects of nanodevices and computational elements. We have synthesized $Ni_{core}$-$Au_{shell}$ attached to DNA. Here Au is conductive and Ni is magnetic. So we have added another more advantage in our material as our material is conductive, optical as well as magnetic also. More over Ni after a gold coating become very stable. This kind of material with both magnetic and conductive properties attached with biomolecule 'DNA' will have great potentiality in medical instrument and computer device.